\documentclass[12pt]{article}
\usepackage[a4paper,top=2cm,bottom=2cm,left=3cm,right=1cm]{geometry}
\usepackage[usenames]{color}
\begin{document}

\title{{\large\bf PROBABILITY REPRESENTATION OF QUANTUM MECHANICS: COMMENTS AND BIBLIOGRAPHY}}
\author{
{\small\bf V. I. Man'ko,}$^1$\ {\small\bf O. V. Pilyavets,}$^1$\ {\small\bf and V. G. Zborovskii}$^2$}
\date{}
\maketitle
\begin{center}
\itshape\small
$^1$P. N. Lebedev Physical Institute, Russian Academy of Sciences\\
Leninskii Pr. 53, Moscow 119991, Russia\\
$^2$Troitsk Institute for Innovation and Fusion Research\\
Pushkovikh 9, Troitsk, Moscow Region 142190, Russia\\
\upshape e-mail:\ \ manko@sci.lebedev.ru\ \ pilyavets@gmail.com\ \ vzborovsky@gmail.com
\ \\
\ \\
\end{center}

\abstract
%\begin{center}\small\bf Abstract\end{center}
%\parshape=3
%0.5cm 17cm 1cm 16.5cm 1cm 16.5cm
{\footnotesize
The probability representation of states in standard quantum mechanics where the
quantum states are associated with fair probability distributions (instead of wave
function or density matrix) is shortly commented and bibliography related to the
probability representation is given}
\footnote{
\parshape=3
0.5cm 17cm 1cm 16.5cm 1cm 16.5cm
The idea to write such preprint for section quant-ph of Los Alamos ArXiv was
suggested by Prof. L.~B.~Okun in connection with seminars on the probability
representation given by one of the authors (V.I.M.) in Institute of Theoretical and
Experimental Physics, Moscow, 2006.}.

\section{Comments}

During last decade a new representation of quantum mechanics called "probability
representation" suggested in [1] was developed. In framework of this representation
the quantum states are described by positive probability distribution functions. The
distribution functions are connected with wave functions or density matrices by known
integral transforms. The probability representation of quantum mechanics, in fact, is
a new formulation of quantum mechanics. There are several formulations of
conventional quantum mechanics described in the review article [D.~F.~Styer, M. S.
Balkin, K.~M.~Becker, M.~R.~Burns, C.~E.~Dudley, S.~T.~Forth, J.~S.~Gaumer,
M.~A.~Kramer, D.~C.~Oertel, L.~H.~Park, M.~T.~Rinkoski, C.~T.~Smith, and
T.~D.~Wotherspoon, \textsl{Am.~J.~Phys.}, \textbf{70}:3 288--297 (2002)] with title
"Nine Formulations Of Quantum Mechanics". Among these formulations there is the
Schr\"odinger one with wave function as basic concept. Also Feynman path integral
formulation of quantum mechanics and Moyal formulation based on evolution equation
for Wigner function of quantum states are discussed in this article. All these nine
formulations are equivalent in the sense of physical results. But the form of
presentation of quantum mechanics is different. The probability representation of
quantum mechanics appeared due to development of quantum tomography (see [J.~Bertrand
and P.~Bertrand, "A Tomographic Approach To Wigner's Function",
\textsl{Found.~Phys.}, \textbf{17}:4 397--405 (1987)] and [K. Vogel and H. Risken,
"Determination Of Quasiprobability Distributions In Terms Of Probability
Distributions For The Rotated Quadrature Phase", \textsl{Phys.~Rev}~A,
\textbf{40}:5-1 2847--2849 (1989)]). The relation of Wigner function to probability
density of homodyne quadrature (optical tomography) was used as a tool to measure
quantum state and the quantum state in the tomographic approach was identified with
the Wigner function. It is worthy to note that the problem of measuring quantum state
(measuring wave function or measuring density matrix) was considered in early days of
quantum mechanics (see, \textit{e.~g.} article by V.~V.~Vladimirskiy: "Recollections
On Meetings With L.~I.~Mandel'shtam", in: \textit{USSR Academy of Sciences,
Departament of General Physics and Astronomy, "Academician L.~I.~Mandel'shtam, 100
Anniversary Of Birth"}, Nauka, Moscow (1979), pp.~207--209 [in~Russian] where
discussions with Mandel'shtam on this topics are reminded). In probability
representation namely tomographic probability distribution is considered as primary
notion of quantum state. This representation is equivalent to all the other ones. It
has some advantage because one can use well known constructions of probability theory
to apply for characteristics of quantum states from the very beginning. 

The text which we present here is not review article on the probability
representation of quantum mechanics. It is, in fact, bibliographic material for
such a review. There are two review articles [A. Vourdas, "Analytic
Representations In Quantum Mechanics", \textsl{J.~Phys.~A:~Math.~Gen.},
\textbf{39}:7 R65--R141 (2006)] and [D.-G. Welsch, W. Vogel, and T.  Opatrn\'y,
"Homodyne Detection And Quantum State Reconstruction", in: E. Wolf (eds),
\textit{Progress in Optics}, North-Holland, Amsterdam (1999), Vol.  XXXIX,
pp.~63--211] as well as some articles of the book [V. V. Dodonov and V. I.
Man'ko (eds.), \textit{Theory of Nonclassical States of Light}, Taylor \&
Francis, London and New York (2003)] where tomographic probability approach is
shortly mentioned. But the complete review on the probability representation is
not available in the literature yet. 

The bibliography on the tomographic probability of quantum mechanics and related to
the probability representation items was collected in connection with writing diploma
thesis on this subject by one of the authors (O.V.P.). The bibliography below
[2--232] has articles with titles ordered according to time of publication. The list
of articles contains both articles on probability representation of continuous
variables and spin degrees of freedom.

Master work~\cite{stud3} and full version of this bibliography containing abstracts
and some more detail information can be downloaded from
http://regulaar.h12.ru/tomographic/

\end{document}